\documentclass[prd,aps, preprintnumbers, showpacs,twocolumn, nofootinbib,superscriptaddress,notitlepage]{revtex4-1}
\usepackage[normalem]{ulem}
\usepackage{epsfig}
\usepackage{amsfonts}
\usepackage{amsmath}
\usepackage{slashed}
\usepackage{graphicx}
\usepackage{color}
\usepackage{mathtools}

\newcommand{\beq}{\begin{eqnarray}}
\newcommand{\eeq}{\end{eqnarray}}
\newcommand{\nn}{\nonumber}

\DeclareRobustCommand{\eq}[1]{Eq.~(\ref{eq:#1})}
\DeclareRobustCommand{\eqs}[2]{Eqs.~(\ref{eq:#1}) and (\ref{eq:#2})}
\DeclareRobustCommand{\sec}[1]{Sec.~\ref{sec:#1}}

\begin{document}

\preprint{MIT-CTP/5074, INT-PUB-19-004}

\title{Matching Quasi Generalized Parton Distributions in the RI/MOM scheme}

\author{Yu-Sheng Liu}
\email{mestelqure@gmail.com}
\affiliation{Tsung-Dao Lee Institute, Shanghai Jiao-Tong University, Shanghai 200240, China}

\author{Wei Wang}
\email{wei.wang@sjtu.edu.cn}
\affiliation{SKLPPC, School of Physics and Astronomy, Shanghai Jiao-Tong University, Shanghai 200240, China}

\author{Ji Xu}
\email{xuji1991@sjtu.edu.cn}
\affiliation{SKLPPC, School of Physics and Astronomy, Shanghai Jiao-Tong University, Shanghai 200240, China}

\author{Qi-An Zhang}
\email{zhangqa@ihep.ac.cn}
\affiliation{Institute of High Energy Physics, Chinese Academy of Science, Beijing 100049, China}
\affiliation{School of Physics, University of Chinese Academy of Sciences, Beijing 100049, China}

\author{Jian-Hui Zhang}
\email{jianhui.zhang@ur.de}
\affiliation{Institut f\"ur Theoretische Physik, Universit\"at Regensburg, D-93040 Regensburg, Germany}
\affiliation{Center of Advanced Quantum Studies, Department of Physics, Beijing Normal University, Beijing 100875, China}

\author{Shuai Zhao}
\email{shuai.zhao@sjtu.edu.cn}
\affiliation{SKLPPC, School of Physics and Astronomy, Shanghai Jiao-Tong University, Shanghai 200240, China}

\author{Yong Zhao}
\email{yzhaoqcd@mit.edu}
\affiliation{Center for Theoretical Physics, Massachusetts Institute of Technology, Cambridge, Massachusetts 02139, USA}

\begin{abstract}
Within the framework of large momentum effective theory (LaMET), generalized parton distributions (GPDs) can be extracted from lattice calculations of quasi-GPDs through a perturbative matching relation, up to power corrections that are suppressed by the hadron momentum. 
In this paper, we focus on isovector quark GPDs, including the unpolarized, longitudinally and transversely polarized cases, and present the one-loop matching that connects the quasi-GPDs renormalized in a regularization-independent momentum subtraction (RI/MOM) scheme to the GPDs in $\overline{\rm MS}$ scheme. We find that the matching coefficient is independent of the momentum transfer squared. As a consequence, the matching for the quasi-GPD with zero skewness is the same as that for the quasi-PDF. Our results provide a crucial input for the determination of quark GPDs from lattice QCD using LaMET. 
\end{abstract}

\maketitle

\section{Introduction}
Understanding the internal structure of nucleons has been an important goal of hadron physics. For many decades, our knowledge on the structure of nucleons has been mainly relying on experimental measurements of their form factors (FFs) and parton distribution functions (PDFs). The FFs describe the spatial distribution of charge and current within the nucleon and can be probed in elastic lepton-nucleon scattering, while the PDFs characterize the longitudinal momentum distribution of quarks and gluons in the nucleon and can be measured in deep-inelastic scattering processes.   

The proposal of generalized parton distributions (GPDs) (for a review, see e.g.~\cite{Ji:2004gf,Diehl:2003ny,Belitsky:2005qn}) provides a novel opportunity to characterize the partonic structure of nucleons. As a generalization of the PDFs to off-forward kinematics, the GPDs contain a wealth of new information on nucleon structure. They naturally encompass the FFs, PDFs as well as the distribution amplitudes (DAs), and offer a description of the correlations between the transverse position and longitudinal momentum of quarks and gluons inside the nucleon, thereby giving access to quark and gluon orbital angular momentum contributions to the nucleon spin. Experimentally, the GPDs can be accessed through hard exclusive processes like deeply virtual Compton scattering or meson production. Much effort has been devoted to measuring such processes at completed and ongoing experiments (HERA~\cite{Adloff:2001cn,Chekanov:2003ya,Aktas:2005ty,Airapetian:2001yk,Airapetian:2011uq,Airapetian:2012mq}, COMPASS~\cite{Gautheron:2010wva}, JLab~\cite{Defurne:2015kxq,Jo:2015ema,Seder:2014cdc,Dudek:2012vr}), and will be continued at planned future facilities such as EIC~\cite{Accardi:2012qut,Aschenauer:2014cki} and EicC~\cite{Chen:2018wyz}. Given the complicated kinematic dependence of GPDs, extracting them from the accumulated experimental data is in general rather difficult, and one usually needs to resort to certain models that allow for an extrapolation to kinematic regions that are not accessible directly~\cite{Bacchetta:2016ccz}.

On the other hand, lattice effort of studying GPDs has been mainly focused on the computation of their moments~\cite{Hagler:2003jd,Gockeler:2003jfa,Schroers:2003mf,Gockeler:2005cj,Hagler:2007xi,Brommel:2007sb,Alexandrou:2011nr}. The full distribution can be reconstructed in principle if all their moments are known. However, the number of moments that are calculable on lattice is very limited, owing to power divergent mixing between different moments operators and increasing stochastic noise for high moments operators.

In the past few years, a new theoretical framework has been developed to circumvent the above difficulties, which is now known as the large momentum effective theory (LaMET)~\cite{Ji:2013dva,Ji:2014gla}. According to LaMET,
the GPDs can be extracted from lattice QCD calculations of appropriately constructed static-operator matrix elements, which are named the quasi-GPDs. The quasi-GPDs are usually hadron-momentum dependent but time independent, and thus can be readily computed on the lattice. After being renormalized nonperturbatively in an appropriate scheme, the renormalized quasi-GPDs can then be matched onto the usual GPDs through a factorization formula accurate up to power corrections that are suppressed by the hadron momentum~\cite{Ji:2015qla,Xiong:2015nua}.

Since LaMET was proposed, a lot of progress has been achieved both with respect to the theoretical understanding of the formalism~\cite{Xiong:2013bka,Ji:2015jwa,Ji:2015qla,Xiong:2015nua,Li:2016amo,Chen:2016utp,Ishikawa:2016znu,Chen:2016fxx,Monahan:2016bvm,Radyushkin:2016hsy,Zhang:2017bzy,Carlson:2017gpk,Ishikawa:2017faj,Xiong:2017jtn,Constantinou:2017sej,Ji:2017oey,Green:2017xeu,Chen:2017mzz,Alexandrou:2017huk,Green:2017xeu,Chen:2017mie,Lin:2017ani,Chen:2017lnm,Rossi:2017muf,Ji:2017rah,Hobbs:2017xtq,Jia:2017uul,Wang:2017qyg,Stewart:2017tvs,Monahan:2017hpu,Wang:2017eel,Izubuchi:2018srq,Xu:2018mpf,Briceno:2018lfj,Xu:2018eii,Jia:2018qee,Spanoudes:2018zya,Rossi:2018zkn,Liu:2018uuj,
Ji:2018waw,Bhattacharya:2018zxi,Radyushkin:2018nbf,Zhang:2018diq,Li:2018tpe,Braun:2018brg} and the direct calculation of PDFs from lattice QCD~\cite{Lin:2014zya,Chen:2016utp,Lin:2017ani,Alexandrou:2015rja,Alexandrou:2016jqi,Alexandrou:2017huk,
Chen:2017mzz,Zhang:2017bzy,Chen:2017gck,Alexandrou:2018pbm,Chen:2018xof,Chen:2018fwa,Alexandrou:2018eet,
Lin:2018qky,Fan:2018dxu,Liu:2018hxv}. The prospects of extracting transverse momentum dependent (TMD) PDFs from lattice with LaMET has been investigated in Refs.~\cite{Ji:2014hxa,Ji:2018hvs,Ebert:2018gzl,Constantinou:2019vyb,Ebert:2019okf}.
In particular, a multiplicative renormalization of both the quark~\cite{Ji:2017oey,
Ishikawa:2017faj,Green:2017xeu} and gluon~\cite{Zhang:2018diq,Li:2018tpe} quasi-PDFs has been established in coordinate space. This allows for a nonperturbative renormalization in the regularization-independent momentum subtraction (RI/MOM) scheme~\cite{Martinelli:1994ty}. For the isovector quark quasi-PDFs, this has been carried out in Refs.~\cite{Chen:2017mzz,Stewart:2017tvs,Chen:2018xof,Lin:2018qky} (see also~\cite{Constantinou:2017sej,Alexandrou:2017huk,Alexandrou:2018pbm}). The relevant hard matching kernel in the same scheme has also been computed up to one loop~\cite{Stewart:2017tvs,Liu:2018uuj,Liu:2018hxv}. Despite limited volumes and relatively coarse lattice spacings, the state-of-the-art nucleon isovector quark PDFs determined from lattice data at the physical point have shown a reasonable agreement~\cite{Chen:2018xof,Lin:2018qky,Alexandrou:2018pbm}  with phenomenological results extracted from the experimental data~\cite{Dulat:2015mca,Ball:2017nwa,Harland-Lang:2014zoa,Nocera:2014gqa,Ethier:2017zbq}. Of course, a careful study of theoretical uncertainties and lattice artifacts is still needed to fully establish the reliability of the results.

As for the GPDs, there have been studies on the perturbative matching of the isovector quark quasi-GPDs~\cite{Ji:2015qla,Xiong:2015nua}, which are free from contributions of disconnected diagrams and mixing with gluon quasi-GPDs. Such studies were performed in a transverse momentum cutoff scheme and therefore not well-suited for the lattice implementation.
In this paper, we reconsider the one-loop matching for isovector quark quasi-GPDs in the RI/MOM scheme.
The results can be used to match the quasi-GPDs calculated in lattice QCD and renormalized in the RI/MOM scheme onto the GPDs in $\overline{\rm MS}$ scheme.

The rest of the paper is organized as follows: In Sec.~\ref{sec:GPD}, we establish our definitions and conventions. In Sec.~\ref{sec:OPE}, we present a rigorous derivation of the factorization formula for the isovector quark quasi-GPD based on operator product expansion (OPE). Section~\ref{sec:renormalization} and~\ref{sec:oneloop} are devoted to the RI/MOM renormalization and matching procedure, respectively. We also explain how to obtain the matching coefficients of DAs from the one-loop results of GPDs in Sec.~\ref{sec:oneloop}. Our summary is given in Sec.~\ref{sec:sum}.

\section{Definitions and conventions}\label{sec:GPD}

The parent function for the quark GPDs, which we call parent-GPD for simplicity, is defined from the Fourier transform of the off-forward matrix element of a light-cone correlator, 
\begin{align}\label{eq:GPD}
&F(\bar\Gamma,x,\xi,t,\mu)\nonumber\\
&=\int \frac{d\zeta^-}{4\pi}e^{-i x\zeta^- P^+}\langle P'',S''|O(\bar\Gamma,\zeta^-)|P',S'\rangle\,,
\end{align}
where $x\in [-1,1]$, the light-cone coordinates $\zeta^\pm=(\zeta^t\pm\zeta^z)/\sqrt{2}$ with $\zeta^\mu=(\zeta^t,\zeta^x,\zeta^y,\zeta^z)$, and the hadron state $|P',S'\rangle$ ($|P'',S''\rangle$) is denoted by its momentum and spin. The parent-GPD is defined in the $\overline{\rm MS}$ scheme and $\mu$ is the renormalization scale.
The kinematic variables are defined as
\begin{align}\label{eq:kinematic_variable}
\Delta\equiv P''-P',\;\; t\equiv \Delta^2,\;\; \xi\equiv -\frac{P''^+-P'^+}{P''^++P'^+}=-\frac{\Delta^+}{2P^+}\,,
\end{align}
where without loss of generality we choose a particular Lorentz frame so that the average momentum
\begin{align}
P^\mu\equiv\frac{P''^\mu+P'^\mu}{2}=(P^t,0,0,P^z)\,,
\end{align}
and only consider the case with $0<\xi<1$.

The light-cone correlator is given by the gauge-invariant nonlocal quark bilinear
\begin{align}
O(\bar\Gamma,\zeta^-)=\bar\psi\left(\frac{\zeta^-}{2}\right)\bar\Gamma\lambda^a W_+\left(\frac{\zeta^-}{2},-\frac{\zeta^-}{2}\right)\psi\left(-\frac{\zeta^-}{2}\right)\,,
\end{align}
{where $\bar\Gamma=\gamma^+$, $\gamma^+\gamma_5$, and $i\sigma^{+\perp}=\gamma^\perp\gamma^+$ correspond to the unpolarized, helicity, and transversity} parent-GPDs, respectively.
$\lambda$ is a Gell-Mann matrix in flavor space, e.g., $\lambda^3$ corresponds to flavor isovector ($u-d$) distribution. The lightlike Wilson line is
\begin{align}
W_+(\zeta_2^-,\zeta_1^-)=P\exp \left[-ig_s\int^{\zeta_2^-}_{\zeta_1^-} A^+(\eta^-)d\eta^-\right]\,.
\end{align}

The GPDs are defined as form factors of the parent-GPD (we follow the convention of Ref. \cite{Diehl:2003ny}),
\begin{widetext}
\begin{align}
F(\bar\Gamma,x,\xi,t,\mu)=\frac{1}{2P^+}\bar{u}(P'',S'')\bigg\{&H(\bar\Gamma,x,\xi,t,\mu)\bar\Gamma+E(\bar\Gamma,x,\xi,t,\mu)\frac{[\slashed \Delta,\bar\Gamma]}{4M}\nonumber\\
&+H'(\bar\Gamma,x,\xi,t,\mu)\frac{P^{[+}\Delta^{\perp]}}{M^2}+E'(\bar\Gamma,x,\xi,t,\mu)\frac{\gamma^{[+}P^{\perp]}}{M}\bigg\}u(P',S')\,,
\end{align}
\end{widetext}
where $[\slashed \Delta,\bar\Gamma]=2i\sigma^{+\mu}\Delta_\mu$, $2\gamma_5\Delta^+$, and $2(\gamma^+\Delta^\perp-\gamma^\perp\Delta^+)$ for $\bar\Gamma=\gamma^+$, $\gamma^+\gamma_5$, and $i\sigma^{+\perp}$, respectively;
$M$ is the hadron mass;
$H$, $E$, $H'$, and $E'$ are the GPDs. Note that $H'$ and $E'$ are nonzero only for transversity GPD.

To calculate the quark GPDs within LaMET, we consider a quark quasi-parent-GPD defined from an equal-time correlator:\footnote{We remind the reader that the tilde notation in GPD community is usually referring to helicity GPDs. In this work, we use tilde notation to specify quasi-GPDs.}
\begin{align}\label{eq:quasiGPD}
&\widetilde F(\Gamma,x,\widetilde{\xi},t,P^z,\widetilde\mu)\nonumber\\
&=\int \frac{dz}{4\pi}e^{i xz P^z}{2P^z\over N}\langle P'',S''|\widetilde O(\Gamma,z)|P',S'\rangle\,,
\end{align}
where $\widetilde{\mu}$ is the renormalization scale in a particular scheme, {and $N$ is a normalization factor that depends on the choice of $\Gamma$. For example, $N=2P^z$ for $\Gamma=\gamma^z$}. The nonlocal quark bilinear \begin{align}\label{eq:tildeO}
\widetilde O(\Gamma,z)=\bar\psi\left(\frac{z}{2}\right)\Gamma\lambda^a W_z\left(\frac{z}{2},-\frac{z}{2}\right)\psi\left(-\frac{z}{2}\right)
\end{align}
is along the $z$ direction with a spacelike Wilson line 
\begin{align}
W_z(z_2,z_1)=P\exp \left[ig_s\int^{z_2}_{z_1} A^z(z')dz'\right]\,.
\end{align}

The kinematic variables are similar to those in Eq.~(\ref{eq:kinematic_variable}) except that the ``quasi" skewness parameter
\begin{align}
\widetilde{\xi}=-\frac{P''^z-P'^z}{P''^z+P'^z}=-\frac{\Delta^z}{2P^z} =  \xi +\mathcal{O}\left({M^2\over P_z^2}\right)\,,
\end{align}
which is equal to $\xi$ up to power corrections. From now on we will replace $\widetilde \xi$ with $\xi$ by assuming that the power corrections are small.

The quasi-GPDs are defined as form factors of the quasi-parent-GPD, 
\begin{widetext}
\begin{align}
\widetilde F(\Gamma,x,\xi,t,P^z,\widetilde\mu)=\frac{1}{N}\bar{u}(P'',S'')\bigg\{&\widetilde H(\Gamma,x,\xi,t,P^z,\widetilde\mu)\Gamma+\widetilde E(\Gamma,x,\xi,t,P^z,\widetilde\mu)\frac{[\slashed \Delta,\Gamma]}{4M}\nonumber\\
&+\widetilde H'(\Gamma,x,\xi,t,\mu)\frac{P^{[z}\Delta^{\perp]}}{M^2}+\widetilde E'(\Gamma,x,\xi,t,\mu)\frac{\gamma^{[z}P^{\perp]}}{M}\bigg\}u(P',S')\,,
\end{align}
\end{widetext}
where $\widetilde H$, $\widetilde E$, $\widetilde H'$, and $\widetilde E'$ are the quasi-GPDs with support $x\in(-\infty,\infty)$. Again, $\widetilde H'$ and $\widetilde E'$ are nonzero only for transversity quasi-GPD. In order to minimize operator mixing on lattice, we choose $\Gamma=\gamma^t$, $\gamma^z\gamma_5$, and $i\sigma^{z\perp}$ for the unpolarized, helicity, and transversity quasi-GPDs~\cite{Constantinou:2017sej,Chen:2017mie}, respectively, {which all correspond to the same normalization factor $N=2P^t$.}

According to LaMET~\cite{Ji:2013dva,Ji:2014gla}, the quasi-GPDs and GPDs are related through a factorization formula. For example,
\begin{widetext}
\begin{align}\label{eq:factorization}
\widetilde{H}(\Gamma,x,\xi,t,P^z,\widetilde{\mu}) = \int_{-1}^1 \frac{dy}{|y|}\, C_{\Gamma}\left(\frac{x}{y},\frac{\xi}{y},{\widetilde{\mu}\over\mu},{\mu\over y P^z}\right) H(\bar{\Gamma},y,\xi,t,\mu)
+\mathcal{O}\left(\frac{M^2}{P_z^2},\frac{t}{P_z^2},\frac{\Lambda_{\rm QCD}^2}{x^2P_z^2}\right)\,,
\end{align}
\end{widetext}
where $M^2/P_z^2$ and $t/P_z^2$ are kinematic power corrections; $\Lambda_{\rm QCD}^2/(x^2P_z^2)$ is the higher-twist correction.
Since the choice of $\Gamma$ corresponds to a unique $\bar{\Gamma}$, we suppress the label $\bar{\Gamma}$ in the matching coefficient $C_{\Gamma}$.
Similar factorization formulas also exist for $\widetilde H'$, $\widetilde E$, and $\widetilde E'$. Equation~(\ref{eq:factorization}) with its explicit form will be rigorously derived in the next section.

\begin{widetext}
\section{Operator product expansion and the factorization formula}\label{sec:OPE}

In this section, we derive the explicit form of the factorization formula for the quasi-GPDs using the OPE of the nonlocal quark bilinear $\widetilde{O}(\Gamma,z)$. The same method has been used for the ``lattice cross section''~\cite{Ma:2017pxb} and quasi-PDF~\cite{Izubuchi:2018srq}, which are both forward matrix elements of a nonlocal gauge-invariant operator. In the case of nonsinglet quasi-PDF, $\widetilde{O}(\gamma^z,z,\mu)$ (e.g., in the $\overline{\rm MS}$ scheme) can be expanded in terms of local gauge-invariant operators in the $|z|\to0$ limit~\cite{Izubuchi:2018srq},
\begin{align}\label{eq:ope-tq}
\widetilde{O}(\gamma^z,z,\mu) = & \sum_{n=0}^\infty \left[C_n ({\mu}^2 z^2)\frac{(-iz)^n}{n!} e_{\mu_1}\cdots e_{\mu_n}O^{\mu_0\mu_1\cdots\mu_n}(\mu) +  \text{higher-twist terms}\right]\,,
\end{align}
where $e^\mu=(0,0,0,1)$, $\mu_0=z$, $C_n=1+O(\alpha_s)$ is the Wilson coefficient, and $O^{\mu_0\mu_1\cdots\mu_n}(\mu)$ is the only allowed renormalized traceless symmetric twist-2 quark operators at leading power in the OPE,
\begin{align}\label{eq:twist-2}
O^{{\mu}_0 {\mu}_1 \ldots {\mu}_n}(\mu) = & Z_{n+1}(\varepsilon,\mu) \bigl[
\bar{\psi} \gamma^{({\mu}_0 } i\overleftrightarrow{D}^{{\mu}_1} \cdots i\overleftrightarrow{D}^{ {\mu}_n)} \psi -
\text{trace} \bigr] \,,
\end{align}
where $\overleftrightarrow{D}=(\overrightarrow{D}-\overleftarrow{D})/2$.
Here $Z_{n+1}^{ij}(\varepsilon,\mu)$ are multiplicative $\overline{\rm MS}$ renormalization factors and $(\mu_0 \cdots \mu_n)$ stands for the symmetrization of these Lorentz indices. Similar technique can be applied to gluon and singlet quark quasi-GPDs by including the corresponding twist-2 operators on the right-hand side of \eq{ope-tq} as well as the mixing between quarks and gluons. Such an extension has been done for the quasi-PDF in Ref. \cite{Wang:2019tgg}.

The multiplicative renormalization shown in \eq{twist-2} is valid for the forward case only, as it is known that in the off-forward case, $O^{{\mu}_0 {\mu}_1 \ldots {\mu}_n}(\mu)$ can mix with other twist-2 operators with overall derivatives $i\partial^\mu$ according to the renormalization group equation~\cite{Braun:2003rp}
\beq \label{eq:rge}
	\mu^2 {d\over d\mu^2}O^{{\mu}_0 {\mu}_1 \ldots {\mu}_n}(\mu) = \sum_{m=0}^{[n/2]}\Gamma_{nm}\left[i\partial^{(\mu_1}\cdots i\partial^{\mu_{2m}} \bar{\psi} \gamma^{\mu_0}i\overleftrightarrow{D}^{\mu_{2m+1}}\cdots i\overleftrightarrow{D}^{\mu_n)}\psi-{\rm trace}\right]\,,
\eeq
where the anomalous dimension $\Gamma$ is an upper triangle matrix.
In off-forward matrix elements, the overall derivative $i\partial^\mu$ contributes a factor of the momentum transfer $\Delta^\mu$. As a result, the OPE in \eq{ope-tq} cannot maintain its form under evolution in $\mu$, so one has to choose the operator bases to be the eigenvectors of \eq{rge} so that each of them is multiplicatively renormalizable.

At leading logarithmic (LL) accuracy, \eq{rge} is diagonalized by the conformal operators~\cite{Efremov:1978rn,Braun:2003rp},
\beq \label{eq:cfm}
	n_{\mu_0}n_{\mu_1}\cdots n_{\mu_n}{\bf O}^{{\mu}_0 {\mu}_1 \ldots {\mu}_n}(\mu) = (in\cdot \partial)^n \bar{\psi} \slashed n\ C_n^{3/2}\left({n\cdot \overrightarrow{D} - n\cdot \overleftarrow{D} \over n\cdot\overrightarrow{\partial}+n\cdot \overleftarrow{\partial}}\right)\psi - {\rm traces}\,,
\eeq
where $n^\mu$ is an arbitrary four vector, and $C_n^{3/2}(\eta)$ is the Gegenbauer polynomial. Beyond LL, the conformal operators start mixing with each other, but \eq{rge} can still be diagonalized with the ``renormalization group improved" conformal operators~\cite{Braun:2003rp,Mueller:1993hg}
\beq \label{eq:newcfm}
	n_{\mu_0}n_{\mu_1}\cdots n_{\mu_n}{\bf O}'^{{\mu}_0 {\mu}_1 \ldots {\mu}_n}(\mu) = \sum_{m=0}^{n}{\cal B}_{nm}(\mu)\left[(in\cdot \partial)^n \bar{\psi} \slashed n\ C_{m}^{3/2}\left({n\cdot \overrightarrow{D} - n\cdot \overleftarrow{D} \over n\cdot\overrightarrow{\partial}+n\cdot \overleftarrow{\partial}}\right)\psi - {\rm traces}\right]\,,
\eeq
where ${\cal B}_{nn}=1$.

As a result, the nonlocal operator $\widetilde{O}(\gamma^z,z,\mu)$ should be generally expanded in terms of these improved conformal operators with modified kinematic factors.

For $\mu_0=\mu_1=\cdots=\mu_n=+$, the off-forward matrix element of the conformal operator ${\bf O}^{{\mu}_0 {\mu}_1 \ldots {\mu}_n}(\mu)$ is given by
\begin{align}\label{eq:ggb}
	\langle P'| {\bf O}^{\overbrace{++\ldots +}^{n+1}}(\mu) |P\rangle  =& \langle P'| (i\partial^+)^n \bar{\psi} \gamma^+ C_n^{3/2}\left({\overrightarrow{D}^+ - \overleftarrow{D}^+\over \overrightarrow{\partial}^++\overleftarrow{\partial}^+}\right)\psi |P\rangle\nn\\
	=& (-\Delta^+)^n (2P^+) \int_{-1}^1 dy\ C_n^{3/2}\left({y\over \xi}\right) F(\gamma^+,y,\xi,t,\mu)\nn\\
	=& (2P^+)^{n+1} \xi^n\int_{-1}^1 dy\ C_n^{3/2}\left({y\over \xi}\right) F(\gamma^+,y,\xi,t,\mu)\,,
\end{align}
which is also known as the Gegenbauer moments.

Using Lorentz covariance, we have for $\mu_0=\mu_1=\cdots=\mu_n=z$,
\begin{align} \label{eq:ggbz}
	\langle P'| {\bf O}^{\overbrace{zz\ldots z}^{n+1}}(\mu) |P\rangle  =&  (2P^z)^{n+1} \xi^n\int_{-1}^1 dy\ C_n^{3/2}\left({y\over \xi}\right) F(\gamma^+,y,\xi,t,\mu)+\mathcal{O}\left({M^2\over P_z^2},{t\over P_z^2}\right)\,,
\end{align}
where $M^2, t\ll P_z^2$, and we have used $\Delta\cdot P=0$. The power corrections originate from the subtracted traces in the kinematic part of the matrix element, and their exact form will be derived in the future.

Based on \eq{ggbz}, we have the leading-twist approximation of the off-forward matrix element of $\widetilde{O}(\gamma^z,z,\mu)$,
\begin{align}\label{eq:ope-twist2}
\langle P'|\widetilde{O}(\gamma^z,z,\mu)|P\rangle  = & 2P^z\sum_{n=0}^\infty C_n ({\mu}^2 z^2){\cal F}_n(-zP^z)\sum_{m=0}^n{\cal B}_{nm}(\mu)\ \xi^n\int_{-1}^1 dy\ C_m^{3/2}\left({y\over \xi}\right) F(\gamma^+,y,\xi,t,\mu) \nn\\
&+ \mathcal{O}\left({M^2\over P_z^2},{t\over P_z^2},z^2\Lambda_{\rm QCD}^2\right)\,,
\end{align}
where ${\cal F}_n(-zP^z)$ are partial wave polynomials whose explicit forms are known in the conformal OPE of current-current correlators for the hadronic light-cone distribution amplitudes~\cite{Braun:2007wv}. The higher-twist terms contribute to $\mathcal O(z^2\Lambda_{\rm QCD}^2)$. 

The polynomiality of $C_n^{3/2}$ allows us to define for $m\le n$,
\beq
	\xi^n C_m^{3/2}\left({y\over \xi}\right) =y^n \left({\xi\over y}\right)^nC_m^{3/2}\left({y\over \xi}\right)  \equiv y^n C'_m\left({\xi\over y}\right)\,,
\eeq
where $C'_m$ is also a polynomial that satisfies
\beq
	C'_m(x) = x^n C_m^{3/2}\left({1\over x}\right)\,.
\eeq

If we define the matching coefficients as
\begin{align}
	\bar{C}_{\gamma^z}\left({x\over \xi}, {y\over \xi}, {\mu \over \xi P^z }\right)=&\int {d(\xi zP^z)\over 2\pi} e^{i{x\over \xi}\xi P^z z} \sum_{n=0}^\infty C_n \left({\mu^2\over (\xi P^z)^2} (\xi zP^z)^2\right){\cal F}_n(-\xi zP^z)\sum_{m=0}^n{\cal B}_{nm}(\mu)\int_{-1}^1 dy\ C_m^{3/2}\left({y\over \xi}\right)\,,\nn\\
	C_{\gamma^z}\left({x\over y}, {\xi\over y}, {\mu\over yP^z}\right)=&\int {d(y zP^z)\over 2\pi} e^{i{x\over y}yP^z z} \sum_{n=0}^\infty C_n \left({\mu^2\over (y P^z)^2} (y zP^z)^2\right){\cal F}_n(-yzP^z)\sum_{m=0}^n{\cal B}_{nm}(\mu)\int_{-1}^1 dy\ C'_m\left({\xi\over y}\right)\,,
\end{align}
then we can Fourier transform \eq{ope-twist2} from $z$ to $xP^z$ to obtain the quasi-GPD and its factorization formula,
\begin{align}
	\widetilde F(\gamma^z,x,\xi,t,P^z,\mu) = & \int_{-1}^1 {dy\over |\xi|}\bar{C}_{\gamma^z}\left({x\over \xi}, {y\over \xi}, {\mu \over \xi P^z }\right)F(\gamma^+,y,\xi,t,\mu) + \mathcal{O}\left({M^2\over P_z^2},{t\over P_z^2},{\Lambda_{\rm QCD}^2\over x^2P_z^2}\right)\,,\label{eq:fact1}\\
	=& \int_{-1}^1 {dy\over |y|}C_{\gamma^z}\left({x\over y}, {\xi\over y}, {\mu \over yP^z }\right)F(\gamma^+,y,\xi,t,\mu) + \mathcal{O}\left({M^2\over P_z^2},{t\over P_z^2},{\Lambda_{\rm QCD}^2\over x^2P_z^2}\right)\,,\label{eq:fact2}
\end{align}
where the second form in \eq{fact2} is postulated in Refs.~\cite{Ji:2015qla,Xiong:2015nua}. Since $xP^z$ is the Fourier conjugate to $z$, the higher-twist contribution of $\mathcal O(z^2\Lambda_{\rm QCD}^2)$ in \eq{ope-twist2} should be of $O(\Lambda_{\rm QCD}^2/(x^2P_z^2))$ in momentum space with an enhancement at small $x$. Such enhancement at small $x$, as well as a $1/(1-x)$ factor, was also found to exist in the power corrections from renormalon ambiguities in the OPE of quasi-PDFs~\cite{Braun:2018brg}. Based on \eqs{fact1}{fact2}, we can infer that the matching coefficients for the quasi-GPDs $\widetilde{H}$, $\widetilde H'$, $\widetilde{E}$, and $\widetilde E'$ must be the same.

For the helicity and transversity quasi-GPDs, $\gamma^z$ in \eq{ope-tq} is replaced by $\gamma^z\gamma^5$ and $i\sigma^{z\perp}$ respectively, and the local twist-two operators $O^{\mu_0\mu_1\cdots\mu_n}$ are also replaced accordingly. This will change the kinematic factors in Eqs.~(\ref{eq:ggb})--(\ref{eq:ope-twist2}), as their tensor structure involves the spin vector of the external state, but it does not affect the form of OPE in \eq{ope-twist2}, nor that of the factorization formulas in \eqs{fact1}{fact2}.

The two matching coefficients in \eqs{fact1}{fact2} are related to each other by
\beq
	C_{\gamma^z}\left({x\over y}, {\xi\over y}, {\mu \over yP^z }\right) = \left|{y\over \xi}\right|\bar{C}_{\gamma^z}\left({x\over \xi}, {y\over \xi}, {\mu \over \xi P^z }\right)\,.
\eeq

The factorization formulas are similar to the evolution equations for the GPD~\cite{Mueller:1998fv,Ji:1996nm}. Notably, at zero skewness $\xi=0$, we have
\beq \label{eq:zeroskew}
	\widetilde F(\gamma^z,x,0,t,P^z,\mu) 
	= \int_{-1}^1 {dy\over |y|}C_{\gamma^z}\left({x\over y}, 0, {\mu \over yP^z }\right)F(\gamma^+,y,0,t,\mu) + \mathcal{O}\left({M^2\over P_z^2},{t\over P_z^2},{\Lambda_{\rm QCD}^2\over x^2P_z^2}\right)\,,
\eeq
where the matching kernel $C_{\gamma^z}(x/y,0,\mu/(yP^z))$ is exactly the same matching coefficient for the $\overline{\rm MS}$ quasi-PDF~\cite{Izubuchi:2018srq}, even when $t\neq0$. Moreover, in the forward limit $\xi\to0$ and $t\to0$, \eq{zeroskew} is exactly the factorization formula for the $\overline{\rm MS}$ quasi-PDF~\cite{Izubuchi:2018srq}.

On the other hand, in the limit $\xi\to1$ and $t\to0$, we obtain the factorization formula for the quasi-DA,
\beq
	\widetilde F(\gamma^z,x,1,t=0,P^z,\mu) =  \int_{-1}^1 dy\ \bar{C}_{\gamma^z}\left(x, y, {\mu \over P^z }\right)F(\gamma^+,y,1,t=0,\mu) + \mathcal{O}\left({M^2\over P_z^2},{\Lambda_{\rm QCD}^2\over x^2P_z^2}\right)\,,
\eeq
whose explicit form has been postulated in Refs.~\cite{Zhang:2017bzy,Xu:2018mpf,Liu:2018tox}.

The same procedure described above also applies to the $\Gamma=\gamma^t$ case. This finishes our derivation of the factorization formula for the isovector quark quasi-GPD, which will enable us to identify the matching coefficients from the one-loop calculation in \sec{oneloop}. 

\end{widetext}

\section{renormalization}\label{sec:renormalization}

Following the strategy in Ref.~\cite{Liu:2018tox}, the UV divergence of the quasi-GPD only depends on the operator $\widetilde O(\Gamma,z)$, not on the external states. We can choose the same renormalization factor as the one for the quasi-PDF~\cite{Stewart:2017tvs,Liu:2018uuj}. For each value of $z$, the RI/MOM renormalization factor $Z$ is calculated nonperturbatively on lattice by imposing the condition that the quantum corrections of the correlator in an off-shell quark state vanish at scales $\{\widetilde{\mu}\}=\{p^2=-\mu_R^2,p^z=p^z_R\}$~\cite{Constantinou:2017sej,Stewart:2017tvs}
\begin{align}\label{eq:Z}
Z(\Gamma,z,a,\mu_R,p^z_R)=\left.\frac{\langle p|\widetilde O(\Gamma,z,a)|p\rangle}{\langle p|\widetilde O(\Gamma,z,a)|p\rangle_{\rm tree}}\right|_{\{\widetilde{\mu}\}}
\end{align}
where $\widetilde O(\Gamma,z,a)$ is the discretized version of $\widetilde O(\Gamma,z)$ on lattice in Eq.~(\ref{eq:tildeO}) with spacing $a$; the bare matrix element $\langle p|\widetilde O(\Gamma,z,a)|p\rangle$ is obtained from the amputated Green's function $\Lambda(\Gamma,z,a,p)$ of $\widetilde O(\Gamma,z,a)$, which is calculated on lattice, with a projection operator ${\cal P}$ for the Dirac matrix
\begin{align}\label{eq:amputated}
\langle p|\widetilde O(\Gamma,z,a)|p\rangle= {\rm Tr}\left[\Lambda(\Gamma,z,a,p){\cal P}\right]\,.
\end{align}

In a systematic calculation of GPD, we start with the bare matrix element of the nonlocal quark bilinear on lattice
\begin{align}
\widetilde{h}(\Gamma,z,\xi,t,P^z,a)={1\over N}\langle P'',S''|\widetilde O(\Gamma,z,a)|P',S'\rangle\,.
\end{align}
After performing RI/MOM renormalization and taking the continuum limit, the renormalized matrix element is
\begin{align}
&\widetilde{h}_R(\Gamma,z,\xi,t,P^z,\mu_R,p^z_R)\nonumber\\
&=\lim_{a\to 0} Z^{-1}(\Gamma,z,a,\mu_R,p^z_R)\widetilde{h}(\Gamma,z,\xi,t,P^z,a)\,,
\end{align}
which is to be Fourier transformed into the $x$-space
\begin{align}
&\widetilde{F}(\Gamma,x,\xi,t,P^z,\mu_R,p^z_R)\nonumber\\
&=P^z \int \frac{dz}{2\pi}e^{i xz P^z}\widetilde{h}_R(\Gamma,z,\xi,t,P^z,\mu_R,p^z_R)\,.
\end{align}
Next, we need to disentangle the terms with different kinematic dependencies to extract quasi-GPDs from $\widetilde F$.
Finally, we match quasi-GPDs in the RI/MOM scheme to GPDs in $\overline{\rm MS}$ scheme according to Eq.~(\ref{eq:factorization}). 
Note that the continuum limit has been taken after the RI/MOM renormalization, we can therefore calculate the matching coefficient in the continuum as the result is regularization independent. For simplicity, we choose dimensional regularization in our calculation. The one-loop result will be presented in the next section.

\begin{widetext}
\section{One-loop matching coefficient}\label{sec:oneloop}

When the hadron momentum $P^z$ is much greater than $M$ and $\Lambda_{\rm QCD}$, the RI/MOM quasi-GPD can be matched onto the $\overline{\rm MS}$ GPD through the factorization formula~\cite{Stewart:2017tvs,Izubuchi:2018srq}
\begin{align}\label{eq:matching2}
\widetilde H(\Gamma,x,\xi,t,P^z,\mu_R,p^z_R)=\int_{-1}^1 \frac{dy}{|y|}\, C_\Gamma\left(\frac{x}{y},\frac{\xi}{y},r,\frac{yP^z}{\mu},\frac{yP^z}{p_R^z}\right) H(\bar\Gamma,y,\xi,t,\mu)+\mathcal{O}\left(\frac{M^2}{P_z^2},{t\over P_z^2},\frac{\Lambda_{\rm QCD}^2}{x^2P_z^2}\right)\,,
\end{align}
\end{widetext}
where $r=\mu_R^2/(p^z_R)^2$. Here we have chosen the explicit form of factorization in \eq{fact2}. To obtain the matching coefficient, we calculate their on-shell massless quark matrix element in perturbation theory by replacing the hadron states in Eqs. (\ref{eq:GPD}) and (\ref{eq:quasiGPD}) with the quark states carrying momentum $p+\Delta/2$ and $p-\Delta/2$ with $p^\mu=(p^t,0,0,p^z)$.

At tree level, the GPDs and quasi-GPDs are 
\begin{align}\label{eq:tree_level}
&H^{(0)}(\bar\Gamma,x,\xi,t)=\widetilde{H}^{(0)}(\Gamma,x,\xi,t,p^z)=\delta(1-x)\, ,\\
&H'^{(0)}=\widetilde{H}'^{(0)}=E^{(0)}=\widetilde{E}^{(0)}=E'^{(0)}=\widetilde{E}'^{(0)}=0\,.
\end{align}

At one-loop order, $\widetilde H^{(1)}$ and $H^{(1)}$ are nonzero and not equal, so their next-to-leading order (NLO) matching kernel is nontrivial;
since $\widetilde H'^{(1)}=H'^{(1)}$, a two-loop calculation is needed to extract the NLO matching kernel;
$\widetilde E^{(1)}$, $E^{(1)}$, $\widetilde E'^{(1)}$, and $E'^{(1)}$ vanish for massless quarks, which agrees with the GPD quark-target model calculation \cite{Meissner:2007rx}. For massive quarks, $\widetilde E^{(1)}=E^{(1)}\neq0$ and $\widetilde E'^{(1)}=E'^{(1)}\neq0$ according to Refs.~\cite{Ji:2015qla,Xiong:2015nua}, so the NLO matching kernel for $\widetilde E^{(1)}$ and $\widetilde E'^{(1)}$
can only be extracted from the two-loop matrix elements in massive quark states. This can be a cross check of the factorization formulas in \eqs{fact1}{fact2}, which, however, is beyond the scope of this work.

In order to combine the ``real" and ``virtual" contributions (defined in Ref.~\cite{Stewart:2017tvs}) in a compact form at one-loop level, we introduce a plus function defined as 
\begin{align}\label{eq:plus}
\int_{-\infty}^\infty dx [h(x)]_+ g(x)&=\int_{-\infty}^\infty dx\, h(x)\big[g(x)-g(1)\big]
\end{align}
with two arbitrary functions $h(x)$ and $g(x)$ which could be piecewise.$h(x)$ can have a single pole at $x=1$, whereas $g(x)$ is regular at $x=1$. By taking the limit $p^t\to p^z$, we obtain the matching kernel for the gauge-invariant bare quasi-GPD and $\overline{\rm MS}$ GPD in a quark,
\begin{widetext}
\begin{align}
C^{(1)}_B\left(\Gamma,x,\xi,\frac{p^z}{\mu},{\mu\over \mu'}\right)=\widetilde H^{(1)}_{B}(\Gamma,x,\xi,t,p^z,\mu',\varepsilon)-H^{(1)}(\bar{\Gamma},x,\xi,t,\mu,\varepsilon)\,,
\end{align}
where the subscript $B$ denotes ``bare" and the ultraviolet (UV) divergence is regulated by dimensional regularization ($D=4-2\varepsilon_{\rm UV}$); the infrared (IR) divergences in $\widetilde H^{(1)}_{B}$ and $H^{(1)}$ are regulated by $t$ and dimensional regularization ($D=4-2\varepsilon_{\rm IR}$), and canceled out in $C^{(1)}_B$; there is still UV divergence remaining due to the virtual contribution for transversity GPD. The results are
\begin{align}\label{eq:bare_matching}
C^{(1)}_B\left(\Gamma,x,\xi,\frac{p^z}{\mu},{\mu\over \mu'}\right)=f_1\left(\Gamma,x,\xi,\frac{p^z}{\mu}\right)_+ +\delta_{\Gamma,i\sigma^{z\perp}}\delta(1-x)\frac{\alpha_s C_F}{4\pi}\left[-\frac{1}{\varepsilon_{\rm UV}}+\ln\left(\frac{\mu^2}{\mu'^2}\right)\right]
\end{align}
where $\delta_{a,b}$ is the Kronecker delta,
\begin{align}
f_1\left(\Gamma,x,\xi,\frac{p^z}{\mu}\right)=\frac{\alpha_s C_F}{2\pi}\left\{
\begin{array}{lc}
G_1(\Gamma,x,\xi)				& x<-\xi\\
G_2(\Gamma,x,\xi,p^z/\mu)		& |x|<\xi\\
G_3(\Gamma,x,\xi,p^z/\mu)		& \xi<x<1\\
-G_1(\Gamma,x,\xi)				& x>1
\end{array}\right.\,,
\end{align}
and
\begin{align}
G_1(\gamma^t,x,\xi)&=G_1(\gamma^z\gamma_5,x,\xi)=-\left[\frac{1}{x-1}-\frac{x}{2\xi}+\frac{1+x}{2(1+\xi)}\right]\ln\frac{x-1}{x+\xi}+(\xi\to-\xi)\,,\\
G_1(i\sigma^{z\perp},x,\xi)&=-\frac{x+\xi}{(x-1)(1+\xi)}\ln\frac{x-1}{x+\xi}+(\xi\to-\xi)\,,\\
G_2\left(\gamma^t,x,\xi,p^z/\mu\right)&=\frac{(x+\xi)(1-x+2\xi)}{2(1-x)\xi(1+\xi)}\left[\ln\frac{4(1-x)^2(x+\xi)(p^z)^2}{(\xi-x)\mu^2}-1\right]+\frac{x+\xi^2}{\xi(1-\xi^2)}\ln\frac{\xi-x}{1-x}\,,\\
G_2\left(\gamma^z\gamma_5,x,\xi,p^z/\mu\right)&=G_2\left(\gamma^t,x,\xi,p^z/\mu\right)+\frac{x+\xi}{\xi(1+\xi)}\,,\\
G_2\left(i\sigma^{z\perp},x,\xi,p^z/\mu\right)&=\frac{x+\xi}{(1-x)(1+\xi)}\left[\ln\frac{4(1-x)^2(x+\xi)(p^z)^2}{(\xi-x)\mu^2}-1\right]+\frac{2\xi}{1-\xi^2}\ln\frac{\xi-x}{1-x}\,,\\
G_3\left(\gamma^t,x,\xi,p^z/\mu\right)&=\frac{1+x^2-2\xi^2}{(1-x)(1-\xi^2)}\left[\ln\frac{4\sqrt{x^2-\xi^2}(1-x)(p^z)^2}{\mu^2}-1\right]+\frac{x+\xi^2}{2\xi(1-\xi^2)}\ln\frac{x+\xi}{x-\xi}\,,\\
G_3\left(\gamma^z\gamma_5,x,\xi,p^z/\mu\right)&=G_3\left(\gamma^t,x,\xi,p^z/\mu\right)+2\frac{1-x}{1-\xi^2}\,,\\
G_3\left(i\sigma^{z\perp},x,\xi,p^z/\mu\right)&=\frac{2(x-\xi^2)}{(1-x)(1-\xi^2)}\left[\ln\frac{4\sqrt{x^2-\xi^2}(1-x)(p^z)^2}{\mu^2}-1\right]+\frac{\xi}{1-\xi^2}\ln\frac{x+\xi}{x-\xi}\,.
\end{align}
\end{widetext}
Some technical details of the calculation are provided in the Appendix. The above calculation has been carried out in momentum space. In principle, the same result can be obtained from calculations in coordinate space and then taking a Fourier transform. For examples in the case of meson DA and nucleon PDF, see Refs.~\cite{Braun:2004bu,Braun:2018brg}. However, as noticed in~\cite{Izubuchi:2018srq,Braun:2018brg}, the step of taking Fourier transform is highly nontrivial. 

We observe that the bare matching coefficients for $\Gamma=\gamma^t$, $\gamma^z\gamma_5$, and $i\sigma^{z\perp}$ reduce to that for the quasi-PDFs~\cite{Izubuchi:2018srq,Liu:2018hxv} when $\xi=0$ even if $t\neq 0$. One can also obtain the bare matching coefficients of DAs \cite{Liu:2018tox} by crossing the external state with the following replacement $\xi\to1/(2y-1)$, $x/\xi\to 2x-1$, and the external momentum {$p^z$ to $p^z/2$} \cite{Ji:2015qla}.

Next we need the counterterm of the quasi-GPD in RI/MOM scheme. As we argued in Sec. \ref{sec:renormalization}, we can use the renormalization factor for the quasi-PDF to renormalize the quasi-GPD, which leads to the one-loop RI/MOM counterterm~\cite{Stewart:2017tvs,Liu:2018uuj}
\begin{widetext}
\begin{align}\label{eq:counterterm}
C^{(1)}_{CT}\left(\Gamma,x,r,\frac{p^z}{p^z_R},\frac{\mu_R}{\mu'}\right)=\left[\left|\frac{p^z}{p^z_R}\right|f_2\left(\Gamma,\frac{p^z}{p^z_R}(x-1)+1,r\right)\right]_+ +\delta_{\Gamma,i\sigma^{z\perp}}\delta(1-x)\frac{\alpha_s C_F}{4\pi}\left[-\frac{1}{\varepsilon_{\rm UV}}+\ln\left(\frac{\mu_R^2}{\mu'^2}\right)\right]\,,
\end{align}
where $r=\mu_R^2/(p^z_R)^2$;
$f_2(\Gamma,x,r)$ is the real part of the off-shell quark matrix element of the quasi-PDF calculated at the subtraction point $\{\widetilde{\mu}\}$; the last term which contains $\delta_{\Gamma,i\sigma^{z\perp}}\delta(1-x)$ is the conversion factor between RI/MOM and $\overline{\rm MS}$ schemes for the local operator $\widetilde{O}(\Gamma,0)$.
We choose Landau gauge, which is convenient for lattice simulation, and project out the coefficient of $\Gamma$ (also known as the minimal projection according to~\cite{Liu:2018uuj}) to obtain $f_2$. The results for different spin structures are \cite{Liu:2018uuj,Liu:2018hxv},
\begin{align}
f_2(\gamma^t,x,r)&=\frac{\alpha_s C_F}{2\pi}\left\{
\begin{array}{lc}
\frac{-3r^2+13rx-8x^2-10rx^2+8x^3}{2(r-1)(x-1)(r-4x+4x^2)}+\frac{-3r+8x-rx-4x^2}{2(r-1)^{3/2}(x-1)}\tan^{-1}\frac{\sqrt{r-1}}{2x-1}  & x>1\\
\frac{-3r+7x-4x^2}{2(r-1)(1-x)}+\frac{3r-8x+rx+4x^2}{2(r-1)^{3/2}(1-x)}\tan^{-1}\sqrt{r-1} & 0<x<1\\
-\frac{-3r^2+13rx-8x^2-10rx^2+8x^3}{2(r-1)(x-1)(r-4x+4x^2)}-\frac{-3r+8x-rx-4x^2}{2(r-1)^{3/2}(x-1)}\tan^{-1}\frac{\sqrt{r-1}}{2x-1} & x<0
\end{array} \right. \,,\\
f_2(\gamma^z\gamma_5,x,r)&=\frac{\alpha_s C_F}{2\pi}\left\{
\begin{array}{lc}
\begin{array}{l}
\frac{3r-(1-2x)^2}{2(r-1)(1-x)}-\frac{4x^2(2-3r+2x+4rx-12x^2+8x^3)}{(r-1)(r-4x+4x^2)^2}+\frac{2-3r+2x^2}{(r-1)^{3/2}(x-1)}\tan^{-1}\frac{\sqrt{r-1}}{2x-1} 
\end{array}
& x>1\\
\frac{1-3r+4x^2}{2(r-1)(1-x)}+\frac{-2+3r-2x^2}{(r-1)^{3/2}(1-x)}\tan^{-1}\sqrt{r-1} & 0<x<1\\
\begin{array}{l}
-\frac{3r-(1-2x)^2}{2(r-1)(1-x)}+\frac{4x^2(2-3r+2x+4rx-12x^2+8x^3)}{(r-1)(r-4x+4x^2)^2}-\frac{2-3r+2x^2}{(r-1)^{3/2}(x-1)}\tan^{-1}\frac{\sqrt{r-1}}{2x-1} 
\end{array}
& x<0
\end{array} \right. \,,\\
f_2(i\sigma^{z\perp},x,r)&=\frac{\alpha_s C_F}{2\pi}\left\{
\begin{array}{lc}
\frac{3}{2(1-x)}+\frac{r-2x}{(r-1)(r-4x+4x^2)}+\frac{-r+2x-rx}{(r-1)^{3/2}(x-1)}\tan^{-1}\frac{\sqrt{r-1}}{2x-1}  & x>1\\
\frac{1-3r+2x}{2(r-1)(1-x)}+\frac{r-2x+rx}{(r-1)^{3/2}(1-x)}\tan^{-1}\sqrt{r-1} & 0<x<1\\
-\frac{3}{2(1-x)}-\frac{r-2x}{(r-1)(r-4x+4x^2)}-\frac{-r+2x-rx}{(r-1)^{3/2}(x-1)}\tan^{-1}\frac{\sqrt{r-1}}{2x-1} & x<0
\end{array} \right. \,.
\end{align}

Finally, combining Eqs. (\ref{eq:bare_matching}) and (\ref{eq:counterterm}), we obtain the one-loop matching coefficient $C_\Gamma$,
\begin{align}
C_\Gamma\left(x,\xi,r,\frac{p^z}{\mu},\frac{p^z}{p^z_R}\right)=\delta(1-x)+C^{(1)}_B\left(\Gamma,x,\xi,\frac{p^z}{\mu},{\mu\over \mu'}\right)-C^{(1)}_{CT}\left(\Gamma,x,r,\frac{p^z}{p^z_R},\frac{\mu_R}{\mu'}\right)+\mathcal{O}(\alpha_s^2)\,;
\end{align}
then making the replacements $x\to x/y$, $\xi\to \xi/y$, and $p^z\to yP^z$~\cite{Stewart:2017tvs,Izubuchi:2018srq}, we obtain $C_\Gamma$ in Eq. (\ref{eq:matching2}),
\begin{align}
C_\Gamma\left({x\over y},{\xi\over y},r,\frac{yP^z}{\mu},\frac{yP^z}{p^z_R}\right)=&\delta\left(1-{x\over y}\right)+\left[f_1\left(\Gamma,{x\over y},{\xi\over y},\frac{yP^z}{\mu}\right)-\left|\frac{y P^z}{p^z_R}\right|f_2\left(\Gamma,\frac{y P^z}{p^z_R}\left({x\over y}-1\right)+1,r\right)\right]_+\nonumber\\
&+\delta_{\Gamma,i\sigma^{z\perp}}\delta\left(1-{x\over y}\right)\frac{\alpha_s C_F}{4\pi}\ln\left(\frac{\mu^2}{\mu_R^2}\right)+\mathcal{O}(\alpha_s^2)\,.
\end{align}
\end{widetext}

\section{summary}\label{sec:sum}

Within the framework of LaMET, we have derived the one-loop matching coefficients that connect the isovector quark quasi-GPDs renormalized in the RI/MOM scheme to GPDs in the $\overline{\rm MS}$ scheme. The calculation was performed for the unpolarized, longitudinally and transversely polarized cases defined with $\Gamma=\gamma^t$, $\gamma^z\gamma_5$, and $i\sigma^{z\perp}$, respectively. We also presented
a rigorous derivation of the factorization formula for isovector quark quasi-GPDs based on OPE. The matching coefficient turns out to be independent of the momentum transfer squared $t$. As a result, for quasi-GPDs with zero skewness the matching coefficient is the same as that for the quasi-PDF. Our results will be used to extract the isovector quark GPDs from lattice calculations of the corresponding quasi-GPDs.
This work can also be extended to gluon and singlet quark quasi-GPDs.

\section*{Acknowledgments}
We thank Vladimir~M. Braun, Yizhuang Liu, Xiangdong Ji, and Yi-Bo Yang for enlightening discussions.
Y.-S. L. is supported by Science and Technology Commission of Shanghai Municipality (Grant No. 16DZ2260200) and National Natural Science Foundation of China (Grant No. 11655002).
W. W., J. X., and S. Z. are supported in part by National Natural Science Foundation of China under Grant No. 11575110, 11655002, 11735010, by Natural Science Foundation of Shanghai under Grants No.~15DZ2272100 and No.~15ZR1423100, Shanghai Key Laboratory for Particle Physics and Cosmology, and  by  MOE  Key Laboratory for Particle Physics, Astrophysics and Cosmology.
Q.-A. Z. is supported by National Natural Science Foundation of China under Grants No. 11621131001 and 11521505.
J.-H. Z. is supported by the SFB/TRR-55 grant ``Hadron Physics from Lattice QCD,'' and a grant from National Science Foundation of China (Grant No.~11405104).
Y. Z. is supported by the U.S. Department of Energy, Office of Science, Office of Nuclear Physics, from DE-SC0011090 and within the framework of the TMD Topical Collaboration. Y. Z. is also partially supported by the Institute for Nuclear Theory at University of Washington during the program INT-18-3 ``Probing Nucleons and Nuclei in High Energy Collisions."

\appendix

\begin{widetext}

\section*{appendix}

\begin{figure}
\includegraphics[width=0.4\columnwidth]{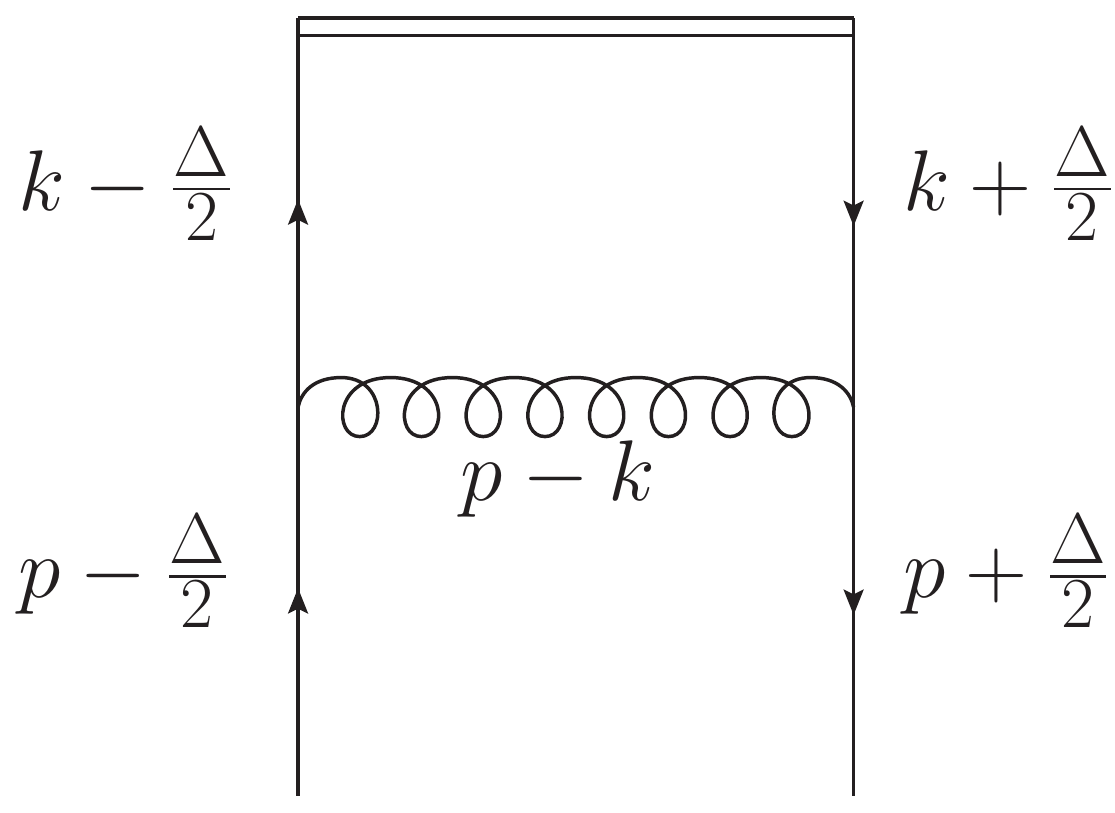}
\caption{One-loop vertex diagram for the quark quasi-GPD. }
\label{fig:diagram}
\end{figure}

In this Appendix, we present some technical details in calculating the following dimensionless integral that arises from the vertex diagram in Fig.~\ref{fig:diagram}:
\begin{align}\label{eq:Idef}
I^N_n(x,\xi) = \int_0^1 dy_1 \int_0^{1-y_1} dy_2 \frac{N(y_1,y_2)}{\left\{y_1 y_2 t'+\left[x-1+(1-\xi)y_1+(1+\xi)y_2\right]^2\right\}^{n+\varepsilon}},
\end{align}
where $N$ is a function of Feynman parameters $y_1$ and $y_2$ and $n$ is the power of the denominator of the integrand; $t'=-t/(p^z)^2$.
In unphysical regions ($x<-\xi$ and $x>1$), the integral has no $1/\varepsilon$ pole so that it can be easily calculated by setting $\varepsilon=0$. However, this is not the case in the Efremov-Radyushkin-Brodsky-Lepage (ERBL), $|x|<\xi$, and Dokshitzer-Gribov-Lipatov-Altarelli-Parisi (DGLAP), $\xi<x<1$, regions where there are IR divergences.

As an example, we evaluate the integral with $N=1$ and $n=3/2$. After integration over $y_2$, the remaining integrand denoted as $F(y_1,\varepsilon)$ contains hypergeometric functions $_2F_1$. We identify the divergent part of $F(y_1,\varepsilon)$ as $A(\varepsilon)/y_1^{1+\varepsilon}$ in the limit of $\varepsilon\to0$, and then separate it out from the integral,
\begin{align}\label{eq:Idecomposition}
I^{1}_{3/2}(x,\xi)=\int_0^1 dy_1 \left[ F(y_1,\varepsilon)-\frac{A(\varepsilon)}{y_1^{1+\varepsilon}} \right]\bigg|_{\varepsilon\rightarrow 0} + \int_0^1 dy_1 \frac{A(\varepsilon)}{y_1^{1+\varepsilon}}\,,
\end{align}
where the first term is convergent so that we can set $\varepsilon=0$ before the integration. We suppress $x$, $\xi$, and $t'$ dependences of $F$ and $A$ for simplicity.

\begin{figure}[htbp]
\includegraphics[width=0.4\columnwidth]{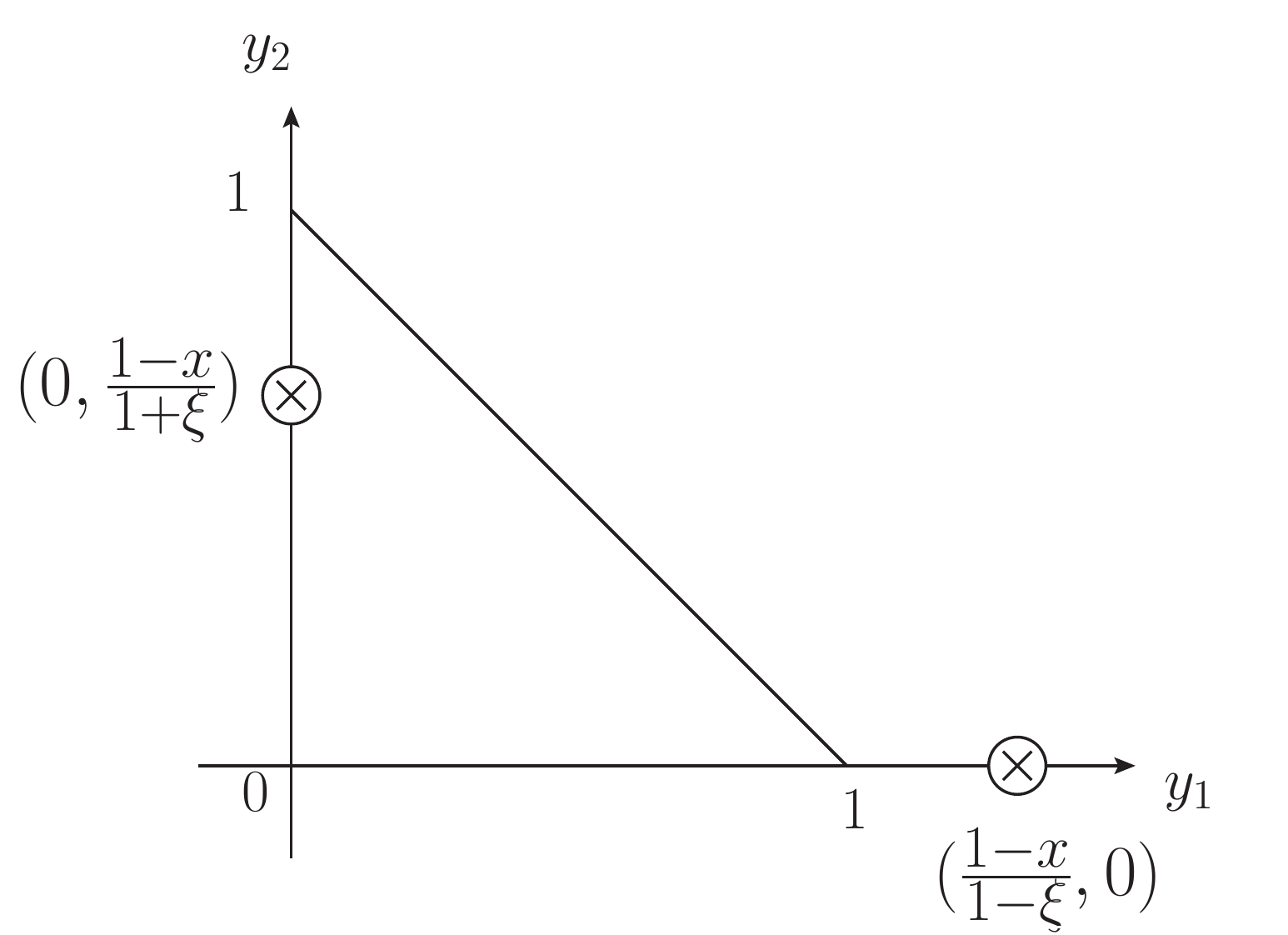}
\includegraphics[width=0.4\columnwidth]{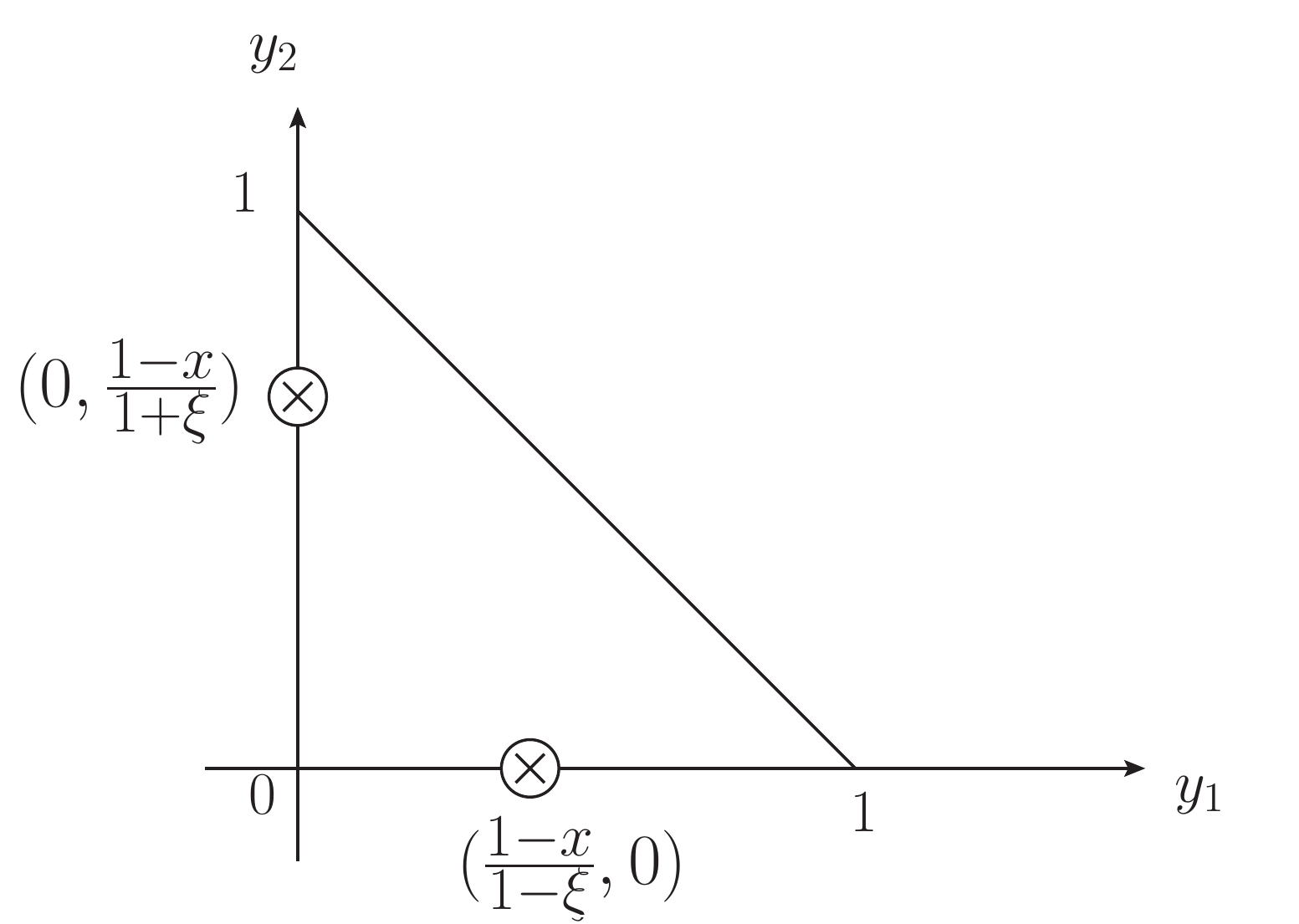}
\caption{Integration in ERBL (left) and DGLAP (right) regions: The singularities are denoted by cross.}
\label{fig:singularities}
\end{figure}

In Fig.~\ref{fig:singularities}, the singularities are shown in integration regions. We use Pfaff transformation to extract the divergent part
\begin{align}
_2F_1(a,b;c;z)=(1-z)^{-a} {}_2F_1(a,c-b;c;\frac{z}{z-1})\, .
\end{align}
We obtain
\begin{align}\label{eq:divergentparts}
A(\varepsilon)=\frac{2}{t'(x-1)}\left\{\begin{array}{lc}
-1+2\varepsilon+\varepsilon \ln{\frac{t'(1-x)}{4(1+\varepsilon)}} &  |x|<\xi  \\
-1+2\varepsilon+\varepsilon \ln{\frac{t'(1-x)}{4(1+\varepsilon)}} -y_1^{1+\varepsilon}\left(1-2\varepsilon-2\varepsilon \ln(1-\xi)\right)\left|y_1-\frac{1-x}{1-\xi}\right|^{-1-2\varepsilon} & \xi<x<1  \end{array} \right. .
\end{align}
Finally, we have
\begin{align}\label{eq:I^1_{3/2}}
I^{1}_{3/2}(x,\xi)= \left\{ \begin{array}{lc}
\frac{1}{2(1-x)(x^2-\xi^2)}   																			& x<-\xi  \\
-\frac{2}{t'(1-x)}\left[ \frac{1}{\varepsilon}-2+\ln\frac{4(\xi-x)(1+\xi)^2}{t'(1-x)^2(x+\xi)} \right]	& |x|<\xi  \\
-\frac{4}{t'(1-x)}\left[ \frac{1}{\varepsilon}-2+\ln\frac{4(1-\xi^2)}{t'(1-x)^2} \right] 				& \xi<x<1 \\
-\frac{1}{2(1-x)(x^2-\xi^2)}   																			& x>1  
\end{array} \right. .
\end{align}

More generally, when calculating the vertex diagram in Fig.(\ref{fig:diagram}), we encounter integrals similar to Eq.(\ref{eq:Idef}) with numerator of the integrand replaced by polynomials of $y_1$ and $y_2$. After integrating out $y_2$, we obtain Appell hypergeometric function $F_1$. In this case, to separate the divergent part, we need Euler transformation
\begin{align}
F_1(\alpha;\beta,\beta';\gamma;x,y)=(1-x)^{-\beta}(1-y)^{-\beta'}F_1(\gamma-\alpha;\beta,\beta';\gamma;\frac{x}{x-1},\frac{y}{y-1})\, .
\end{align}

In the following, we list integrals used in our calculation:
\begin{enumerate}
\item In unphysical region $x<-\xi$, there is no divergence.
\begin{align}
I^{1}_{1/2}&=\frac{1-x}{1-\xi^2}\ln\left|\frac{1-x}{x-\xi}\right|+\frac{x+\xi}{2\xi(1+\xi)}\ln\left|\frac{x-\xi}{x+\xi}\right|\,,\\
I^{y_1}_{3/2}&=-\frac{1}{4\xi(1-\xi)(x-\xi)}+\frac{1}{2(1+\xi)(1-\xi)^2}\ln\left|\frac{1-x}{x-\xi}\right|+\frac{1}{8\xi^2(1+\xi)}\ln\left|\frac{x-\xi}{x+\xi}\right|\,,\\
I^{y_2}_{3/2}&=I^{y_1}_{3/2}\Big|_{\xi\rightarrow-\xi}\,,\\
I^{y_1^2}_{3/2}&=\frac{x-3x\xi+2\xi^2}{4\xi^2(x-\xi)(1-\xi)^2}+\frac{1-x}{(1+\xi)(1-\xi)^3}\ln\left|\frac{1-x}{x-\xi}\right|+\frac{\xi+x}{8\xi^3(1+\xi)}\ln\left|\frac{x-\xi}{x+\xi}\right|\,,\\
I^{y_2^2}_{3/2}&=I^{y_1^2}_{3/2}\Big|_{\xi\rightarrow-\xi}\,,\\
I^{y_1y_2}_{3/2}&=-\frac{1}{4\xi^2(1-\xi^2)}+\frac{1-x}{2(1-\xi^2)^2}\ln\left|\frac{1-x}{x-\xi}\right|-\frac{x+2x\xi+\xi^2}{8\xi^3(1+\xi)^2}\ln\left|\frac{x-\xi}{x+\xi}\right|.
\end{align}

\item In ERBL region $|x|<\xi$, $I^{y_2}_{3/2}$ and $I^{y_2^2}_{3/2}$ are divergent.
\begin{align}
I^{1}_{1/2}&=\frac{x+\xi}{2\xi(1+\xi)}\ln\frac{16\xi^2}{t'}+\frac{1-x}{1-\xi^2}\ln\frac{1+\xi}{2\xi}\,,\\
I^{y_1}_{3/2}&=-\frac{2}{t'(1-\xi)}\ln\frac{(\xi-x)(1+\xi)}{2\xi(1-x)}\,,\\
I^{y_2}_{3/2}&=-\frac{2}{t'(1+\xi)}\left[\frac{1}{\varepsilon}-2+\ln\frac{8\xi(1+\xi)}{t'(1-x)(x+\xi)}\right]\,,\\
I^{y_1^2}_{3/2}&=-\frac{x+\xi}{t'\xi(1-\xi)}+\frac{2(1-x)}{t'(1-\xi)^2}\ln\frac{2\xi(1-x)}{(1+\xi)(\xi-x)}\,,\\
I^{y_2^2}_{3/2}&=-\frac{(1-\xi)(x+\xi)}{t'\xi(1+\xi)^2}-\frac{2(1-x)}{t'(1+\xi)^2}\left[\frac{1}{\varepsilon}-2+\ln\frac{8\xi(1+\xi)}{t'(1-x)(x+\xi)}\right]\,,\\
I^{y_1y_2}_{3/2}&=\frac{x+\xi}{t'\xi(1+\xi)}\,.
\end{align}

\item In DGLAP region $\xi<x<1$, $I^{y_1}_{3/2}$, $I^{y_2}_{3/2}$, $I^{y_1^2}_{3/2}$, and $I^{y_2^2}_{3/2}$ are divergent.
\begin{align}
I^{1}_{1/2}&=\frac{x-\xi^2}{2\xi(1-\xi^2)}\ln\frac{x+\xi}{x-\xi}+\frac{1-x}{1-\xi^2}\ln\frac{4(1-\xi^2)\sqrt{x^2-\xi^2}}{t'(1-x)}\,,\\
I^{y_1}_{3/2}&=\frac{1-x}{2(1-\xi)}I^{1}_{3/2}\,,\\
I^{y_2}_{3/2}&=\frac{1-x}{2(1+\xi)}I^{1}_{3/2}\,,\\
I^{y_1^2}_{3/2}&=\frac{(1-x)^2}{2(1-\xi)^2}I^{1}_{3/2}-\frac{2(1-x)}{t'(1-\xi)^2}\,,\\
I^{y_2^2}_{3/2}&=\frac{(1-x)^2}{2(1+\xi)^2}I^{1}_{3/2}-\frac{2(1-x)}{t'(1+\xi)^2}\,,\\
I^{y_1y_2}_{3/2}&=\frac{2(1-x)}{t'(1-\xi^2)}\,.
\end{align}

\item In unphysical region $x>1$, the integrals are the same as functions in another unphysical region but with an overall minus sign, $I^{P(y_1,y_2)}_n(x>1)=-I^{P(y_1,y_2)}_n(x<-\xi)$. See Eq. (\ref{eq:I^1_{3/2}}) for an example.
\end{enumerate}

\end{widetext}

\bibliographystyle{apsrev4-1}
\bibliography{bibliography}

\end{document}